\documentclass[aps,prc,preprint,tightenlines,showpacs,%
amssymb,byrevtex,nofootinbib,endfloats]{revtex4}

\usepackage{epsfig,bm,dcolumn}

\begin{document}

\preprint{hep-ph/0307286}

%%%%%%%%%%%%%%%%%%%%% Title %%%%%%%%%%%%%%%%%%%%%%

\title{Higher meson resonances in $\bm{\rho \to \pi^0 \pi^0 \gamma}$
and $\bm{\omega \to \pi^0 \pi^0 \gamma}$}

%%%%%%%%%%%%%%%%%%%% Authors %%%%%%%%%%%%%%%%%%%%%
%%%%%%%%%%%%%%%%%%%% Addresses %%%%%%%%%%%%%%%%%%%%%

\author{Yongseok Oh}%
\email{yoh@phya.yonsei.ac.kr}

\author{Hungchong Kim}%
\email{hung@phya.yonsei.ac.kr}

\affiliation{Institute of Physics and Applied Physics,
Yonsei University, Seoul 120-749, Korea}

\date{\today}

%%%%%%%%%%%%%%%%%%%% Abstract %%%%%%%%%%%%%%%%%%%%%

\begin{abstract}

The role of higher meson resonances with spin 1 and 2 is investigated
quantitatively in the decay processes of $\rho \to \pi^0\pi^0 \gamma$
and $\omega \to \pi^0 \pi^0 \gamma$.
Among the higher resonances, we find that the $f_2(1270)$ tensor meson
can give a nontrivial contribution especially to the
$\omega \to \pi^0 \pi^0 \gamma$ decay process.
When the $f_2$ contribution is combined with the processes involving the
vector and scalar meson intermediate states, 
a good agreement with the recent measurements is achieved for both decays.
The effect of the $f_2(1270)$ is found to be sizable at the intermediate
photon energies and may be verified by precise measurements of the
recoil photon spectrum of the $\omega \to \pi^0 \pi^0 \gamma$ decay.
The dependence of the decay widths on various models for the $\rho$-$\omega$
mixing in the literature is also investigated.

\end{abstract}

\pacs{13.20.Jf, 13.40.Hq}

\maketitle

\section{Introduction}

The radiative decays of vector mesons into two neutral pseudoscalar
mesons are expected to be useful to resolve the puzzles of the scalar
mesons such as $\sigma(500)$, $f_0(980)$, and $a_0(980)$.
Among the possible decays of $V \to PP\gamma$, where
$V=\rho,\omega,\phi$ and $P = \pi^0, \eta, K^0, \bar{K}^0$,
the SND experiments \cite{SND02,SND00} recently reported the first
measurement for the $\rho \to \pi^0 \pi^0 \gamma$ decay and confirmed the
GAMS experiment \cite{GAMS94} for the $\omega \to \pi^0 \pi^0 \gamma$
decay with higher accuracy.
The measured branching ratios are
\begin{eqnarray}
\mbox{BR}(\rho \to \pi^0 \pi^0 \gamma) =
(4.1 \stackrel{+1.0}{\mbox{\scriptsize $-0.9$}} \pm 0.3)
\times 10^{-5}, \nonumber \\
\mbox{BR}(\omega \to \pi^0 \pi^0 \gamma) =
(6.6 \stackrel{+1.4}{\mbox{\scriptsize $-0.8$}} \pm 0.6)
\times 10^{-5}.
\end{eqnarray}

Theoretically, since the pioneering works of Singer \cite{Singer62,Singer63}
based on the vector meson dominance, various models have been developed
so far.
The decay widths were estimated in the scheme of current algebra
\cite{Renard69} or with SU(3) chiral Lagrangian \cite{FO90,BGP92}.
For $\rho,\omega \to \pi^0 \pi^0 \gamma$ decays at the tree level, one
considers the two diagrams shown in Fig.~\ref{fig:types}, i.e., either
$V \to (M) \pi^0 \to (\pi^0 \gamma) \pi^0$ or $V \to (M) \gamma \to
(\pi^0 \pi^0) \gamma$, where $M$ is the intermediate meson.
Since the lowest pseudoscalar mesons are not allowed in the intermediate
state, an important contribution is expected to come from the vector meson
channel, i.e., when $M = \omega$ or $\rho$.
However, the estimate~\cite{BGP92} based on the vector meson dominance, 
$\rho \to (\omega) \pi^0 \to \gamma \pi^0 \pi^0$ and 
$\omega \to (\rho) \pi^0 \to \gamma \pi^0 \pi^0$,
gives too small decay widths for both decays compared with the experimental
data.
The momentum-dependent decay width of the intermediate vector meson and the 
$\rho$-$\omega$ mixing are found to improve the situation only 
slightly~\cite{GS01}.
In particular, for the $\rho \to \pi^0 \pi^0 \gamma$ decay, these effects
are almost negligible.

In addition to the tree level decay process, final state interactions
can contribute to the decay amplitudes.
Specifically the radiative decays of vector mesons into two pseudoscalar
mesons offer a possibility to study the final state interactions in a given
hadronic channel and they are expected to shed light on the investigation
of the scalar mesons.
In Ref.~\cite{BGP92b}, Bramon {\it et al.\/} applied chiral perturbation
theory to estimate the contribution from pion and kaon loops in
$V \to \pi^0 \pi^0 \gamma$ at the one-loop level.
They found that, when the pion loop is included, $\Gamma(\rho \to \pi^0
\pi^0 \gamma)$ becomes doubled, which is however still lower than the
experimental value, and $\Gamma(\omega \to \pi^0\pi^0 \gamma)$ remains
almost unaltered.
Therefore, the discrepancy with the experimental data persists and further
improvements are necessary.

For $\Gamma(\rho \to \pi^0 \pi^0 \gamma)$, the intermediate $\sigma(500)$
meson contribution could be important.
The first attempt in this direction was made in the framework of unitarized
chiral loops \cite{MHOT99,PHO02}, where the $\sigma$ meson is dynamically
generated from the unitary resummation of the pion loops.
This approach, which corresponds to the process $\rho \to \sigma \gamma \to
\pi^0\pi^0\gamma$, could lead to a good agreement with the recent data for
the $\rho \to \pi^0\pi^0\gamma$ decay width.

Alternatively, one may introduce the $\sigma$ meson explicitly in an
effective Lagrangian approach.
%But the $\sigma$ meson contribution from the direct $\rho\sigma\gamma$
%coupling as initially suggested by Refs.~\cite{GY00,GY01a} overestimates
The initial calculation of the $\sigma$ meson contribution
using direct $\rho\sigma\gamma$ coupling~\cite{GY00,GY01a} however 
overestimates
the measured $\Gamma(\rho \to \pi^0\pi^0\gamma)$ by two orders of
magnitude.
This unrealistic result is later ascribed~\cite{BELN01,GSY03} to the
large and momentum-independent $\rho\sigma\gamma$ coupling
($g_{\rho\sigma\gamma} \approx 2.2 \sim 3.2$)
determined from the analyses of $\rho$ meson photoproduction near threshold
\cite{FS96,OTL00} or by QCD sum rules \cite{GY01,AOS02}.
For the magnitude of the $\rho\sigma\gamma$ coupling, the SND experiment
extracted $\Gamma(\rho \to \sigma \gamma)$ in a {\em model-dependent\/} way
\cite{SND02}, which leads to a small coupling $g_{\rho\gamma\sigma} \approx 
0.25$ \cite{OL03}.
Moreover, the recent study on $\rho$ photoproduction at low energies
\cite{OL03} shows that the $\sigma$ exchange model with large
$g_{\rho\gamma\sigma}$ can be successfully
replaced by the $f_2$ tensor meson and two-pion exchanges.
This example also suggests that the $\rho\sigma\gamma$ coupling is not
necessarily large to explain the $\rho$ photoproduction data near threshold.
For the momentum-dependence of the $\rho\sigma\gamma$ coupling, Bramon
{\em et al.\/} \cite{BELN01} used the linear sigma model, where the direct
$\rho\sigma\gamma$ coupling is not allowed.
Instead, the $\sigma$ couples to $\rho\gamma$ through chiral loops,
which then gives an effective (and momentum-dependent) $\rho\sigma\gamma$
coupling~\cite{BELN01}.
The obtained results show strong dependence on the mass and the
decay width of the $\sigma$ meson \cite{GSY03,RT03,BE03} so that the
$\rho \to \pi^0 \pi^0 \gamma$ decay process may constrain such parameters.

Although the $\sigma$ meson contribution could give a reasonable
description for the $\rho \to \pi^0\pi^0\gamma$ decay width, the
$\omega \to \pi^0\pi^0\gamma$ process still waits for an improvement.
This is because the chiral loop and the intermediate $\sigma$ meson give only
negligible corrections to $\omega \to \pi^0\pi^0\gamma$.
The $\sigma$ meson couples to $\omega\gamma$ through a charged-pion loop but
the $\omega \pi^+ \pi^-$ vertex violates the $G$ parity.
The charged kaon loops can contribute but its magnitude is estimated to be
three orders of magnitude smaller than that from the vector meson dominance
model \cite{BGP92b}.
Another approach is therefore developed by allowing the direct
$\omega\sigma\gamma$ coupling~\cite{GY00,GKY03,RT03}, which however needs
further theoretical justifications in the viewpoint of the linear sigma
model as the coupling occurs only through the chiral loops.
Thus, it would be interesting to look for additional mechanisms that
can mostly contribute to the $\omega \to \pi^0\pi^0\gamma$ decay.

In this paper, we explore the role driven by the other mesons with spin
1 and 2 that have not been accounted for in the processes,
$\rho \to \pi^0\pi^0\gamma$ and $\omega \to \pi^0\pi^0\gamma$.
In particular, we estimate the contribution from the $f_2(1270)$ spin-2
tensor meson.
The meson states with higher masses have been neglected so far because
they have been expected to give small contributions~\cite{BELN01,PHO02,BE03}.
This could be true as far as the magnitude of the $f_2(1270)$ meson
contribution is concerned.
However, through the strong interference with the other decay amplitudes,
the $f_2(1270)$ meson can give sizable contributions
especially for the $\omega \to \pi^0 \pi^0\gamma$ decay.

This paper is organized as follows.
In Section II, the general features of the $V \to \pi^0 \pi^0 \gamma$
decay at the tree level are discussed with kinematics.
Various decay amplitudes are then discussed in Section III.
The results are presented and compared in Section IV.
Section V contains a summary.

\section{General Features and Kinematics}

\begin{figure}[t]
\centering
\epsfig{file=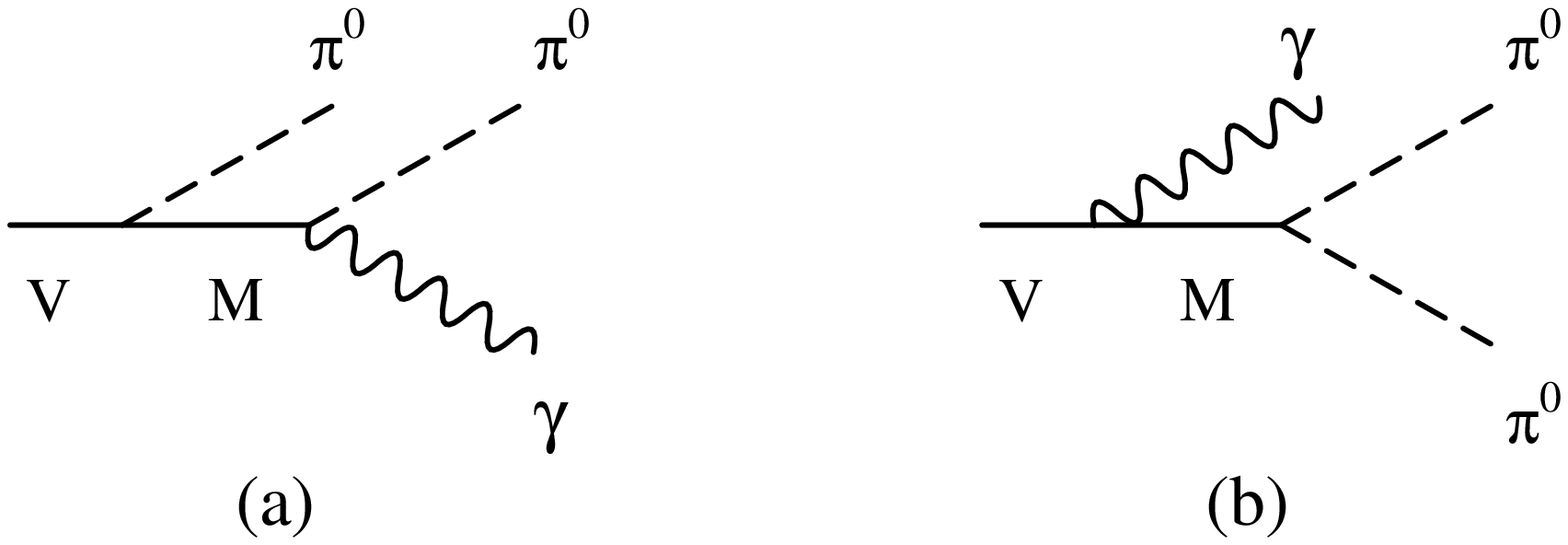, width=0.7\hsize}
\caption{Tree diagrams for $V \to \pi^0 \pi^0 \gamma$ decay
($V=\rho,\omega$), where the the crossed diagrams are understood in (a)
and $M$ includes the possible intermediate meson states.}
\label{fig:types}
\end{figure}

We start with the possible tree diagrams which can contribute to the $V
\to \pi^0 \pi^0 \gamma$ decay as shown in Fig.~\ref{fig:types}.
Figure~\ref{fig:types}(a) includes the crossed diagram which
has the same intermediate meson state in the decay of our interest
($V=\rho,\omega$).
The intermediate meson $M$ now includes $V = (\rho,\omega)$,
$a=(a_1,a_2)$, and $f=(f_1,f_2)$.
For the intermediate scalar mesons we just follow the work of
Refs.~\cite{BELN01,BE03} that will be summarized in Section III.
The quantum numbers of the intermediate mesons are
$I^G(J^{PC}) = 0^+(1^{++})$ for $f_1$,
$0^+(2^{++})$ for $f_2$, $1^-(1^{++})$ for $a_1$, $1^-(2^{++})$ for
$a_2$, $1^+(1^{--})$ for $\rho$, and $0^-(1^{--})$ for $\omega$.
Because of the quantum numbers, only limited intermediate mesons
are allowed for the decay process of Fig.~\ref{fig:types}.

For the $\rho \to \pi^0 \pi^0 \gamma$ decay, only $M=\omega$ is allowed for
the diagram of Fig.~\ref{fig:types}(a) since the $\rho\rho\pi$ and $a\pi\pi$
vertices are forbidden by $G$ parity and the $\rho f \pi$ vertex violates
$G$ and $C$ parities.
In addition, the $\rho\omega\gamma$, $\rho\rho\gamma$, $\rho a \gamma$,
$a\pi\pi$, and $\omega\pi\pi$ couplings are forbidden by $C$ and/or $G$
parities.
Therefore only $M=f$ is allowed for Fig.~\ref{fig:types}(b).

For the $\omega \to \pi^0 \pi^0 \gamma$ decay, the $\omega\omega\pi$
$\omega f \pi$, $f \pi \gamma$, $a\pi\gamma$, $\omega a \pi$,
$\rho\omega\gamma$, $\omega\omega\gamma$, $\omega\pi\pi$, and $a\pi\pi$
couplings are forbidden by the same reasons.
Therefore, only $M=\rho$ and $M=f$ are allowed for Fig.~\ref{fig:types}(a)
and Fig.~\ref{fig:types}(b), respectively.

Figure~\ref{fig:types}(a) with the intermediate vector meson state,
i.e., $M=\omega$ ($\rho$) for $\rho$ ($\omega$) decay, is the vector meson
dominance model that has been considered in the previous studies.
But Fig.~\ref{fig:types}(b) with $M=f$ has not been considered so far
\cite{PHO02}.
We also notice that the decay mode of the $f_1(1285)$, $f_1(1420)$, and 
$f_2(2010)$ into two pions have not been observed yet. 
The $f_2'(1525)$ contribution through Fig.~\ref{fig:types}(b) is suppressed
because of its negligible branching ratio to the $\pi\pi$ decay.
However, the $f_2(1270)$ meson is special since it decays mostly into
$\pi\pi$ and its decay into $\gamma\gamma$ is known \cite{PDG02}, which
constrains the $f_2 V \gamma$ vertices through vector meson dominance.
Thus, the $f_2(1270)$ meson contribution through Fig.~\ref{fig:types}(b)
is expected to be nontrivial.

In our calculation, we denote the four-momenta of the initial vector meson,
outgoing photon, and outgoing pions by $p_\mu$, $k_\mu$, $p_{1\mu}$,
and $p_{2\mu}$, respectively.
The decay width for $V \to \pi^0 \pi^0 \gamma$ can be calculated from
\begin{equation}
\Gamma(V \to \pi^0 \pi^0 \gamma)
= \frac12 \int_{k_{\rm min}}^{k_{\rm max}} dk \int_{E_{\pi, \rm
min}}^{E_{\pi, \rm max}} dE_\pi \frac{d^2\Gamma}{dk dE_\pi},
\label{Gam1}
\end{equation}
where $k$ is the outgoing photon energy and $E_\pi$ is the energy of an
outgoing pion.
The factor $1/2$ accounts for two identical particles in the
final state and
\begin{equation}
 \frac{d^2\Gamma}{dk dE_\pi} =
\frac{1}{(2\pi)^3} \frac{1}{24 M_V} \sum_{\rm spins} |\mathcal{M}|^2,
\end{equation}
where $M_V^{}$ is the initial vector meson mass and the sum runs over
the polarizations of the initial and final vector particles.
We write the decay amplitude $\mathcal{M}$ as
\begin{equation}
\mathcal{M} = \varepsilon^{*(\gamma)}_\mu M^{\mu\nu} \varepsilon^{(V)}_\nu,
\end{equation}
where $\varepsilon^{*(\gamma)}_\mu $ ($\varepsilon^{(V)}_\nu$) is 
the photon (vector meson) polarization vector.
Then Eq.~(\ref{Gam1}) can be rewritten as
\begin{equation}
\Gamma(V \to \pi^0\pi^0\gamma) = \frac{1}{384 \pi^3 M_V} \int_0^{k_{\rm
max}} dk F_V(k),
\end{equation}
where
\begin{equation}
F_V(k) =
\int_{E_{\pi, \rm min}}^{E_{\pi, \rm max}} dE_\pi \sum_{\rm spins}
|\mathcal{M}|^2,
\end{equation}
and
\begin{eqnarray}
k_{\rm max} &=& \frac{1}{2M_V} \left( M_V^2 - 4 M_\pi^2 \right), \nonumber
\\
E_{\pi, \rm max/min} &=& \frac12 \left\{ (M_V - k) \pm k \sqrt{
\frac{2(k_{\rm max}-k)}{M_V-2k}} \right\}.
\end{eqnarray}

\section{Decay Amplitudes}

\subsection{Intermediate vector meson}

We first consider the $M=V'$ case in Fig.~\ref{fig:types}(a).
For this purpose, we use the following effective Lagrangian,
\begin{eqnarray}
\mathcal{L}_{\omega\rho\pi} &=& g_{\omega\rho\pi}^{}
\epsilon^{\mu\nu\alpha\beta} \partial_\mu \omega_\nu \partial_\alpha
\rho_\beta^0 \pi^0, \nonumber \\
\mathcal{L}_{V\pi\gamma} &=& \frac{eg_{V\pi\gamma}^{}}{M_V}
\epsilon^{\mu\nu\alpha\beta} \partial_\mu V_\nu \partial_\alpha A_\beta
\pi^0,
\end{eqnarray}
where $V_\mu = \omega_\mu, \rho^0_\mu$ and $A_\mu$ is the photon field.
The coupling constants $g_{V\pi\gamma}$ are determined from the
experimental data for $\Gamma(V \to \pi^0 \gamma)$ \cite{PDG02}, which gives
\begin{equation}
g_{\rho\pi\gamma} = 0.756, \qquad g_{\omega\pi\gamma} = 1.843.
\end{equation}
The $g_{\omega\rho\pi}$ cannot be determined directly from experiments,
but theoretical estimates lie between 11 GeV$^{-1}$ and 15
GeV$^{-1}$. Here we use $g_{\omega\rho\pi} = 14.9$ GeV$^{-1}$ following
the hidden gauge approach of Refs.~\cite{BGP92,BKY88}.

The direct and crossed diagrams then give 
\begin{eqnarray}
M^{V'}_{\mu\nu} &=& \frac{eg_{\omega\rho\pi}g_{V'\pi\gamma}}{M_{V'}}
\epsilon_{\mu\rho\sigma\tau} \epsilon_{\nu\alpha\beta\gamma} p^\gamma
k^\tau g^{\alpha\rho}
\nonumber \\ && \mbox{} \times
\left( \frac{P^\beta P^\sigma}{P^2 - M_{V'}^2 + i M_{V'}\Gamma_{V'}}
+ \frac{Q^\beta Q^\sigma}{Q^2 - M_{V'}^2 + i M_{V'}\Gamma_{V'}}
\right),
\label{VDM}
\end{eqnarray}
where $P= p-p_1$ and $Q = k+p_1$.
Here $V' = \omega$ for the $\rho$ decay and $V'=\rho$ for the $\omega$
decay.
As discussed in Ref. \cite{GS01}, introducing the momentum dependence in 
the decay width $\Gamma_{V'}$ increases the total decay width by about
10\% for the case of $\omega \to \pi^0 \pi^0 \gamma$ decay.
Following Refs.~\cite{GS01,OPTW95-97,GS68}, we use
\begin{equation}
\Gamma_V(q^2) = \Gamma_V \left( \frac{q^2 - 4 M_\pi^2}{M_V^2 - 4
M_\pi^2} \right)^{3/2} \frac{M_V^{}}{\sqrt{q^2}} \ \theta(q^2-4M_\pi^2),
\end{equation}
where $\theta(x)$ is the step function.
We use $\Gamma_\rho = 150.7$ MeV and $\Gamma_\omega = 8.44$ MeV
\cite{PDG02}.
This effect is however negligible for $\rho \to \pi^0 \pi^0 \gamma$
since $\Gamma_\omega$ is small.

\subsection{Intermediate tensor meson}

In this work, we use an effective Lagrangian approach to introduce
the $f_2(1270)$ resonance.
The effective Lagrangian should take into account the spin-2 structure
of the $f_2$ meson.%
\footnote{For a dynamical generation of the $f_2(1270)$ from chiral
Lagrangian, see, e.g., Ref.~\cite{DP02}.}
The effective Lagrangian for the $f_2\pi\pi$ interaction reads
\cite{PSMM73}
\begin{equation}
\mathcal{L}_{f\pi\pi} = - \frac{2 G_{f\pi\pi}}{M_f} \partial_\mu
\bm{\pi} \cdot \partial^\mu \bm{\pi} f^{\mu\nu},
\label{fpipi}
\end{equation}
where $f^{\mu\nu}$ is the $f_2$ meson field of mass $M_f$.
The coupling constant is determined from the experimental data for
$\Gamma(f_2 \to \pi\pi)$ \cite{PDG02} as 
\begin{equation}
\frac{G_{f\pi\pi}^2}{4\pi} \approx 2.64,
\label{eq:Gfpp}
\end{equation}
which gives $G_{f\pi\pi} \approx 5.76$.
This coupling is used to fix the universal coupling constant of the
tensor meson, which then determines the $f_2$-nucleon and $f_2$--vector-meson
coupling constants through the tensor meson dominance
\cite{Renn70,Renn71,Raman71a,OL03}.

For the  $f_2^{}V\gamma$ vertex, we make use of the tensor
meson dominance.
The most general form for the gauge-invariant $f_2V\gamma$ vertex reads
\cite{Renn71}
\begin{equation}
\langle \gamma(k) f_2 | V(k') \rangle = \frac{1}{M_f}
\epsilon^\kappa \epsilon'^\lambda f^{\mu\nu} A_{\kappa \lambda
\mu\nu}^{fV\gamma}(k,k'),
\label{FVV}
\end{equation}
where
\begin{eqnarray}
A^{fV\gamma}_{\kappa\lambda\mu\nu} (k,k') &=&
-\frac{f^{}_{fV\gamma}}{M_f^3} \left[ g_{\kappa \lambda} (k\cdot k')
- k'_\kappa k_\lambda \right] (k+k')_\mu (k+k')_\nu 
\nonumber \\ && \mbox{}
+ g_{fV\gamma}^{} [ g_{\kappa\lambda} (k+k')_\mu (k+k')_\nu
- g_{\lambda\mu} k'_\kappa (k+k')_\nu
- g_{\lambda\nu} k'_\kappa (k+k')_\mu
\nonumber \\ && \mbox{} \qquad\qquad
- g_{\kappa\mu} k_\lambda (k+k')_\nu
- g_{\kappa\nu} k_\lambda (k+k')_\mu
\nonumber \\ && \mbox{} \qquad\qquad
+2 k \cdot k' ( g_{\kappa\mu} g_{\lambda\nu} + g_{\kappa\nu}
g_{\lambda\mu}) ].
\label{AFVV}
\end{eqnarray}

Since there is no experimental information for the decay of $f_2 \to V
\gamma$, we have to rely on some assumptions for the coupling constants.
First the tensor meson dominance and vector meson dominances lead
to~\cite{Renn71}
\begin{eqnarray}
f_{fV\gamma}^{} = 0, \qquad
g_{fV\gamma}^{} = \frac{e}{f_V} G_{fVV},
\label{coups}
\end{eqnarray}
where $G_{fVV}$ is the $f_2 VV$ coupling constant and $f_V^{}$ the
vector meson decay constant ($f_\rho = 5.33$, $f_\omega = 15.2$).
Since tensor meson dominance gives $G_{fVV} = G_{f\pi\pi}$, the coupling
constants can be fixed from the above relations.
However, the coupling constants obtained in this way overestimate the
observed decay width of $f_2 (1270) \to \gamma\gamma$ by a factor of
3-4.%
\footnote{Note that the vertex (\ref{FVV}) can also be applied to
the $f_2\to\gamma\gamma$ decay. 
If we apply the vector meson dominance once more to the relation
(\ref{coups}), we have $f_{f\gamma\gamma}^{} = 0$ and
$g_{f\gamma\gamma}^{} = e^2 \left( {1}/{f_\rho^2} +
{1}/{f_\omega^2} \right) G_{fVV}$.}
This may be related to the corrections to the tensor and vector meson
dominances and experimental measurements of $\Gamma(f_2 \to V \gamma)$
should clarify this issue.
In this paper, we either work with the relation $G_{fVV} = G_{f\pi\pi}$ or
fix $G_{fVV}$ from the data for $\Gamma(f_2 \to \gamma\gamma)$.
The latter procedure yields $G_{fVV} = 3.12$.
The details on the tensor meson dominance can be found, e.g.,
in Ref.~\cite{OL03}.

It is now straightforward to obtain the decay amplitude with the
intermediate $f_2$ meson as
\begin{equation}
M^{f}_{\mu\nu} = \frac{G_{f\pi\pi}}{M_f^2} \frac{1}{R^2 - M_f^2 + i M_f
\Gamma_f} (2p_1-p+k)_\alpha (2p_1-p+k)_\beta P^{\alpha\beta,\rho\sigma}
A_{\mu\nu\rho\sigma},
\end{equation}
where $R = p-k$ and $P^{\alpha\beta;\rho\sigma}$ comes from the tensor
meson propagator,
\begin{equation}
P^{\mu\nu;\rho\sigma} = \frac12 \left( \bar{g}^{\mu\rho}
\bar{g}^{\nu\sigma} + \bar{g}^{\mu\sigma} \bar{g}^{\nu\rho} \right) -
\frac13 \bar{g}^{\mu\nu} \bar{g}^{\rho\sigma},
\end{equation}
with
\begin{equation}
\bar{g}_{\mu\nu} = -g_{\mu\nu} + \frac{R_\mu R_\nu}{M_f^2}.
\label{gbar}
\end{equation}
Note that there is no ambiguity on the relative phases between the
couplings in the amplitude above, since they are fixed by tensor and
vector meson dominances.

For the momentum dependence of the $f_2$ decay width $\Gamma_f$,
we use the form%
\footnote{For a quark model description for the $f_2$ decay width $\Gamma_f$,
see, for example, Ref. \cite{GI85}.}
\begin{equation}
\Gamma_f(q^2) = \Gamma_f \left( \frac{q^2 - 4 M_\pi^2}{M_f^2 - 4
M_\pi^2} \right)^{5/2} \frac{M_f^{2}}{q^2} \ \theta(q^2-4M_\pi^2),
\end{equation}
which is motivated by the Lagrangian (\ref{fpipi}).
The total decay width of the $f_2$ is $\Gamma_f = 185.7$ MeV \cite{PDG02}.

\subsection{Chiral loops and scalar meson}

\begin{figure}
\centering
\epsfig{file=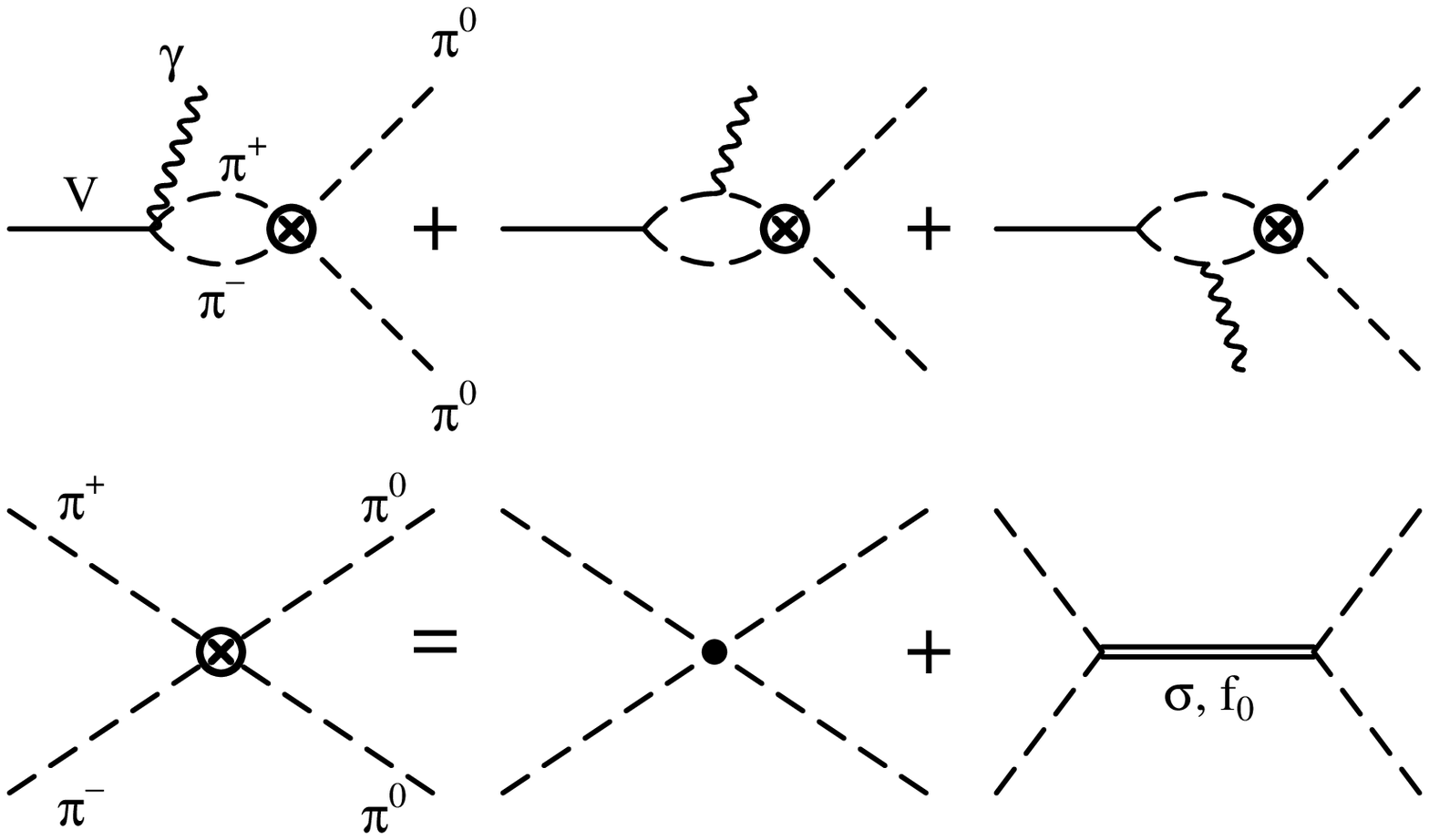, width=0.7\hsize}
\caption{One-loop diagrams for $V \to \pi^0 \pi^0 \gamma$.}
\label{fig:loop}
\end{figure}

The contributions from the chiral loops and scalar mesons were studied
using chiral perturbation theory \cite{BGP92b}, unitarized chiral
perturbation theory \cite{MHOT99,PHO02}, and linear sigma model
\cite{BELN01,BE03}.
Here we follow the work of Bramon {\it et al.\/} \cite{BELN01,BE03}
based on the linear sigma model.
The one-loop Feynman diagrams for $V \to \pi^0 \pi^0 \gamma$ are 
shown in Fig.~\ref{fig:loop}. 
The loop diagrams calculated analytically are found to be finite
and we have \cite{BGP92b,BELN01}
\begin{equation}
M^{\chi\mbox{-}\sigma}_{\mu\nu} =
- \frac{e g_\rho}{\sqrt2 \pi^2 M_{\pi^+}^2} \left( p \cdot k g_{\mu\nu}
- p_\mu k_\nu \right) L (m_{\pi^0 \pi^0}^2)
\mathcal{A}(\pi^+ \pi^- \to \pi^0 \pi^0),
\end{equation}
where $m_{\pi^0 \pi^0}^2 = M_V^2 - 2 M_V k$ and
\begin{equation}
L (m_{\pi^0 \pi^0}^2) = \frac{1}{2(a-b)} - \frac{2}{(a-b)^2} \left[
f(1/b) - f(1/a) \right] + \frac{a}{(a-b)^2} \left[ g(1/b) - g(1/a)
\right],
\end{equation}
with $a = M_V^2/M_{\pi^+}^2$, $b = (M_V^2 - 2 M_V k)/M_{\pi^+}^2$, and
$g_\rho = 4.27$.
The loop integrals $f(z)$ and $g(z)$ are defined in Ref. \cite{BELN01}
and will not be repeated here.
Finally the amplitude $\mathcal{A}(\pi^+ \pi^- \to \pi^0 \pi^0)$ reads
\begin{equation}
\mathcal{A}(\pi^+ \pi^- \to \pi^0 \pi^0) = \frac{m_{\pi^0 \pi^0}^2 -
M_\pi^2}{f_\pi^2} F_\sigma(m_{\pi^0\pi^0}^2),
\end{equation}
which reduces to the chiral loop results \cite{BGP92b} in the limit of
large $\sigma$ meson mass.  
Neglecting the $f_0(980)$ contribution,%
\footnote{It has been claimed that the contribution from the $f_0(980)$ is
suppressed mainly because of its large mass \cite{BELN01}.}
we can write \cite{BE03}
\begin{equation}
F_\sigma(s) = \frac{-M_\sigma^2 + \kappa M_\pi^2}{D_\sigma(s)},
\end{equation}
where $D_\sigma(s) = s - M_\sigma^2 + i M_\sigma \Gamma_\sigma$. 
Here we set $M_\sigma = 500$ MeV and $\kappa = 1$, which corresponds to 
$\Gamma_\sigma \approx 300$ MeV \cite{BE03}.
The $\sigma$ decay width also has the momentum dependence as
\cite{GKY03}
\begin{equation}
\Gamma_\sigma(q^2) = \Gamma_\sigma \left( \frac{q^2 - 4 M_\pi^2}{M_\sigma^2
- 4 M_\pi^2} \right)^{1/2} \frac{M_\sigma^{2}}{q^2} \ \theta(q^2-4M_\pi^2).
\end{equation}
We will not consider the kaon loops as their contribution is suppressed
\cite{BGP92b}.
Furthermore, since $\omega\pi\pi$ coupling violates the $G$ parity, the pion
loop does not contribute to $\omega \to \pi^0 \pi^0 \gamma$ in the good
isospin limit.
The observed $\omega \to \pi^+ \pi^-$ decay can contribute only through
the $\rho$-$\omega$ mixing, which will be discussed later.
Therefore, in the scheme of linear sigma model, the pion loops and
scalar mesons can contribute only to the $\rho \to \pi^0 \pi^0 \gamma$
decay unless the $\rho$-$\omega$ mixing is included.

\subsection{$\rho$-$\omega$ mixing}

The $\rho$-$\omega$ mixing in the $V \to \pi^0\pi^0\gamma$ decays
was first considered by Guetta and Singer \cite{GS01} and found to
increase the $\omega \to \pi^0 \pi^0 \gamma$ decay width by about 5\%.
The mixing is responsible for the nonvanishing branching ratio of 
the $G$-violating $\omega \to \pi^+ \pi^-$ decay and may be described by
adding a term $\mathcal{M}_{\rho\omega}^2 \omega_\mu \rho^\mu$ to the
effective Lagrangian \cite{OPTW95-97}.
With the $\rho$-$\omega$ mixing, the physical vector meson states are
\begin{equation}
\rho = \rho^{(I=1)} + \epsilon \omega^{(I=0)}, \qquad
\omega = \omega^{(I=0)} + \epsilon \rho^{(I=1)},
\end{equation}
where
\begin{equation}
\epsilon = \frac{\mathcal{M}_{\rho\omega}^2}{M_\omega^2 - M_\rho^2 + i
M_\rho \Gamma_\rho - i M_\omega \Gamma_\omega}\ .
\end{equation}

\begin{figure}
\centering
\epsfig{file=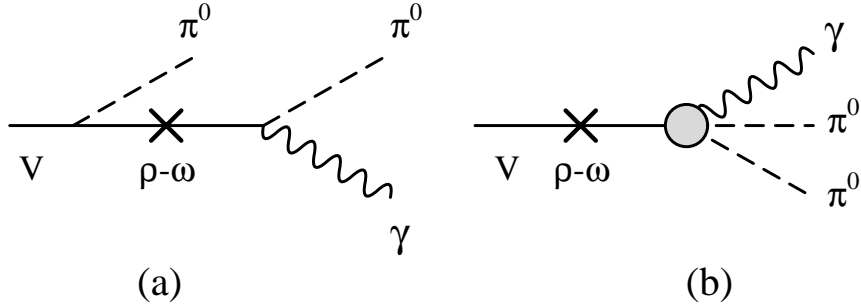, width=0.7\hsize}
\caption{$\rho$-$\omega$ mixing in $V \to \pi^0 \pi^0 \gamma$ decay. The
blob in (b) includes the whole decay amplitude for
$V' \to \pi^0 \pi^0 \gamma$.}
\label{fig:mix}
\end{figure}

In the case of $\omega \to \pi^0 \pi^0 \gamma$ decay, 
the amplitude including the mixing reads \cite{GS01}
\begin{equation}
M^{\rm mix}_{\mu\nu}(\omega \to \pi^0 \pi^0 \gamma) =
\tilde{M}_{\mu\nu} (\omega \to \pi^0 \pi^0 \gamma)
+ \epsilon M_{\mu\nu}(\rho \to \pi^0 \pi^0 \gamma),
\label{eq:mix}
\end{equation}
where $\tilde{M}_{\mu\nu} (\omega \to \pi^0 \pi^0 \gamma)$ is the decay
amplitude with the intermediate $\rho$ meson (\ref{VDM}) 
whose propagator is replaced by 
\begin{eqnarray}
\frac{1}{P^2 - M_\rho^2 + i M_\rho \Gamma_\rho} &\to&
\frac{1}{P^2 - M_\rho^2 + i M_\rho \Gamma_\rho}
\nonumber \\ && \mbox{} +
\frac{M_\rho}{M_\omega} \frac{g_{\omega\pi\gamma}}{g_{\rho\pi\gamma}}
\frac{\mathcal{M}_{\rho\omega}^2}{(P^2 - M_\rho^2 + i M_\rho
\Gamma_\rho)(P^2 - M_\omega^2 + i M_\omega \Gamma_\omega)}.
\end{eqnarray}
The second term of Eq.~(\ref{eq:mix}) multiplied by $\epsilon$
includes the whole decay amplitude for $\rho \to \pi^0 \pi^0 \gamma$.
The mixing accounts for the $G$-violating $\omega \to \pi^+ \pi^-$ process
so that the $\sigma$ meson implicitly participates in the
$\omega \to \pi^0 \pi^0 \gamma$ decay.
The mixing parameter $\mathcal{M}_{\rho\omega}^2$ will be discussed in
the next Section.
Note that in order to make a consistent description with the
$\rho \to \pi^0\pi^0 \gamma$ decay, we do not consider the direct
$\omega\sigma\gamma$ coupling, which is different from the models of
Refs.~\cite{GY00,GKY03,RT03}.

The $\rho$-$\omega$ mixing effect should also be considered for the
$\rho \to \pi^0 \pi^0 \gamma$ decay, which will lead to coupled
equations for the $\rho$ and $\omega$ decays.
However, since $g_{\omega\pi\gamma}>g_{\rho\pi\gamma}$ and
$\mathcal{M}(\rho \to \pi^0 \pi^0 \gamma)$ is larger than $\mathcal{M}(\omega
\to \pi^0 \pi^0 \gamma)$, we expect that 
%its effect 
the mixing can be safely neglected
for the case of $\rho \to \pi^0 \pi^0 \gamma$ decay as implied
by the $\rho$-$\omega$ mixing formula (\ref{eq:mix}).
In this work, therefore, we consider the $\rho$-$\omega$ mixing effect in
$\omega \to \pi^0 \pi^0 \gamma$ only.

\section{Results}

\begin{figure}
\centering
\epsfig{file=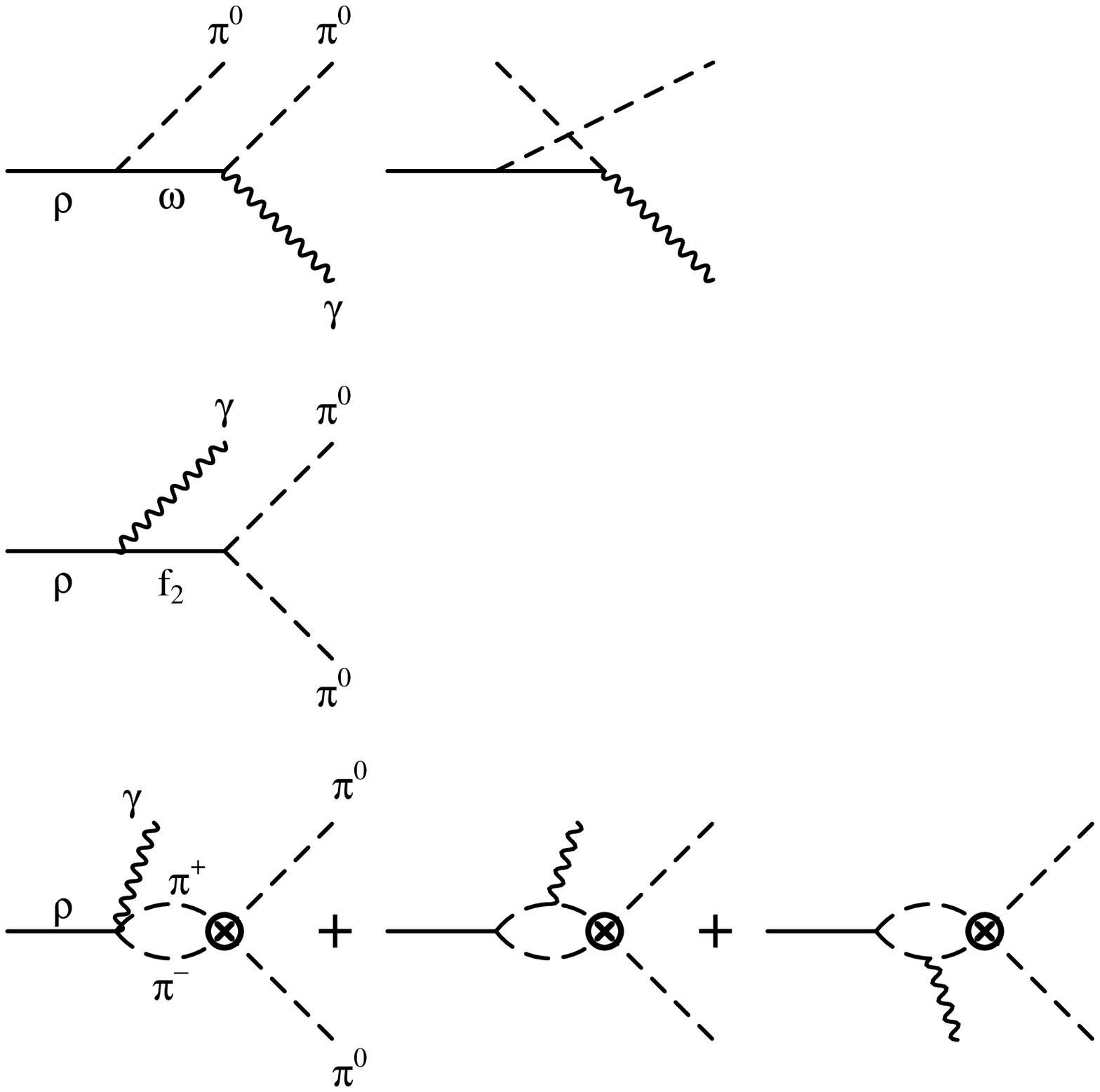, width=0.7\hsize}
\caption{Diagrams for the decay of $\rho \to \pi^0 \pi^0 \gamma$. The
crossed-circle vertex is given in Fig.~\ref{fig:loop}.}
\label{fig:rho}
\end{figure}

\begin{figure}
\centering
\epsfig{file=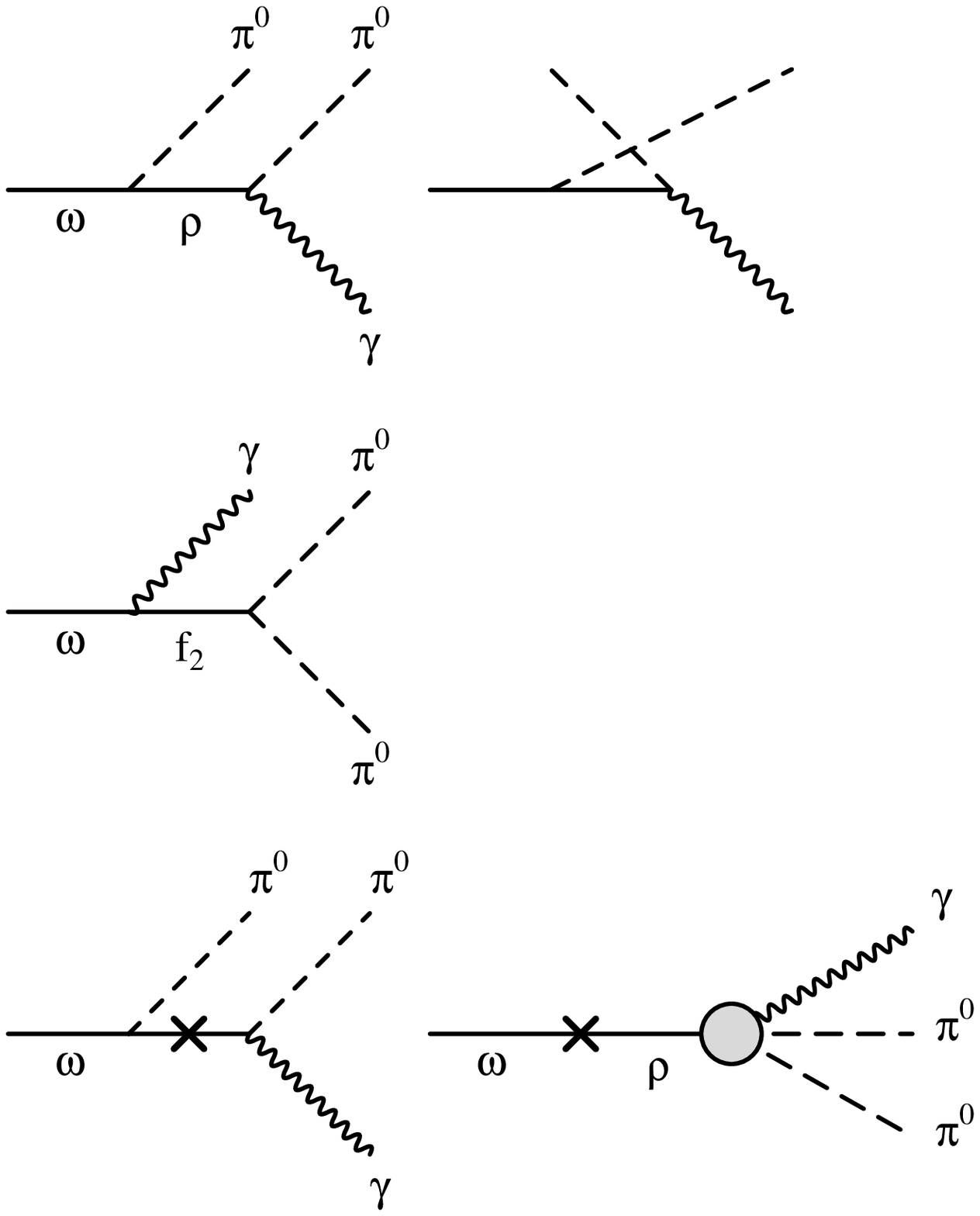, width=0.6\hsize}
\caption{Diagrams for the decay of $\omega \to \pi^0 \pi^0 \gamma$. The cross
denotes the $\rho$-$\omega$ mixing as in Fig.~\ref{fig:mix}.}
\label{fig:omega}
\end{figure}

We now combine the results of Section III to complete our model for
the $\rho,\omega \to \pi^0 \pi^0 \gamma$ decays.
We will consider the intermediate vector meson, $f_2$ tensor meson,
chiral loop and $\sigma$ meson, and $\rho$-$\omega$ mixing.
The $\rho$-$\omega$ mixing will be considered only in the 
$\omega \to \pi^0 \pi^0 \gamma$ decay since its effect is expected to be
suppressed in $\rho \to \pi^0 \pi^0 \gamma$.
The chiral loop and $\sigma$ meson will be considered for both
decays.
Since there is no direct coupling of $V\sigma\gamma$, the $\sigma$ meson
couples to the $\rho\gamma$ only through the pion loop and to the
$\omega\gamma$ only through the $\rho$-$\omega$ mixing.
This is different from the model of Ref. \cite{GKY03}, where the direct
$\omega\sigma\gamma$ coupling is allowed to explain the observed decay
width for $\omega \to \pi^0 \pi^0 \gamma$.
Our models for $\rho \to \pi^0 \pi^0 \gamma$ and $\omega \to \pi^0
\pi^0 \gamma$ decays are shown in Figs.~\ref{fig:rho} and \ref{fig:omega}.

In order to see the role of the $f_2$ meson, we work with three models.
Model (A) includes the intermediate vector meson only, model (B)
contains the intermediate vector meson plus $\rho$-$\omega$ mixing, and
model (C) is constructed by adding the chiral loop and $\sigma$ meson
contributions to model (B). 
In each model, we investigate how the intermediate $f_2$ meson 
modifies the predicted decay widths.
As emphasized in Refs.~\cite{GSY03,BE03}, the $\sigma$ meson contribution
is strongly dependent on the parameters of the $\sigma$ meson.
Therefore, careful analyses on the experimental data for the decay width
and its photon energy distribution are required to constrain the $\sigma$
meson parameters from the $\rho \to \pi^0 \pi^0 \gamma$ decay
\cite{SND02}.
The decay width for $\omega \to \pi^0 \pi^0 \gamma$ is, however,
insensitive to the $\sigma$ meson contribution.
As discussed before, this is because the $\sigma$ meson can couples to
$\omega\gamma$ only through the $\rho$-$\omega$ mixing.
For the $\rho$-$\omega$ mixing parameter, we use \cite{OPTW95-97}
\begin{equation}
\mathcal{M}_{\rho\omega}^2 = - (3.8 \pm 0.4) \times 10^3 \mbox{ MeV}^2.
\end{equation}

Our results for the $\rho \to \pi^0 \pi^0 \gamma$ and $\omega \to \pi^0
\pi^0 \gamma$ decays are summarized in Table~\ref{tab:res}.
Without the $f_2$ meson contribution, we can see the $\sigma$ meson
contribution is crucial in reproducing the experimental branching ratio
of $\rho \to \pi^0 \pi^0 \gamma$.
But, as expected, this does not change the result for 
$\omega \to \pi^0 \pi^0 \gamma$ much.
Then, in each model, we include the contribution from the $f_2$
meson.
For the coupling $g_{fV\gamma}^{}$, we first use the tensor meson
dominance relation $G_{fVV} = G_{f\pi\pi}$ in Eq.~(\ref{coups}).
As anticipated from the previous studies \cite{BELN01,PHO02}, the magnitude
of the intermediate $f_2$ meson alone is suppressed, namely 
we have 70 eV for $\Gamma(\rho \to \pi^0 \pi^0 \gamma)$ 
and 10 eV for $\Gamma(\omega \to \pi^0 \pi^0 \gamma)$.
However, due to the strong interference with the other decay amplitudes,
its contribution to the final result is noticeable, i.e., about
5\% to $\Gamma(\rho \to \pi^0 \pi^0 \gamma)$
and 20\% to $\Gamma(\omega \to \pi^0 \pi^0 \gamma)$.

\begin{table}
\centering
\begin{ruledtabular}
\begin{tabular}{ccccc}
 &
\multicolumn{2}{c}{$\mbox{BR}(\rho \to \pi^0\pi^0\gamma) \times 10^5$} & 
\multicolumn{2}{c}{$\mbox{BR}(\omega \to \pi^0\pi^0 \gamma) \times 10^5$} \\
\raisebox{1.5ex}{Model}
    &  without $f_2$ & with $f_2$
    &  without $f_2$ & with $f_2$
\\ \hline
(A) &   1.10    & 1.37 (1.23)   & 5.02   &  5.92 (5.47)    \\
(B) &   1.10    & 1.37 (1.23)   & 5.08   &  5.96 (5.52)    \\
(C) &   4.09    & 4.27 (4.18)   & 5.15   &  6.36 (5.88)    \\ 
\hline
Expt. &
\multicolumn{2}{c}{$4.1 \stackrel{+1.0}{\mbox{\scriptsize
$-0.9$}} \pm 0.3$ } &
\multicolumn{2}{c}{$6.6 \stackrel{+1.4}{\mbox{\scriptsize $-0.8$}} \pm
0.6$ } \\
\end{tabular}
\end{ruledtabular}
\caption{Calculated branching ratios, $\mbox{BR}(\rho \to \pi^0\pi^0\gamma)$
and $\mbox{BR}(\omega \to \pi^0\pi^0 \gamma)$. Model (A)
includes the intermediate vector mesons only and model (B) includes
the vector mesons and the $\rho$-$\omega$ mixing. Model (C) includes the
chiral loop and the $\sigma$ meson contribution in addition to the
mechanisms of model (B). In each model,
the predictions without and with the $f_2$ meson are given with
$G_{fVV} = 5.763$.
The values in the parentheses are obtained with $G_{fVV} = 3.12$. The
experimental values are from Ref. \cite{SND02}.}
\label{tab:res}
\end{table}

\begin{figure}
\centering
\epsfig{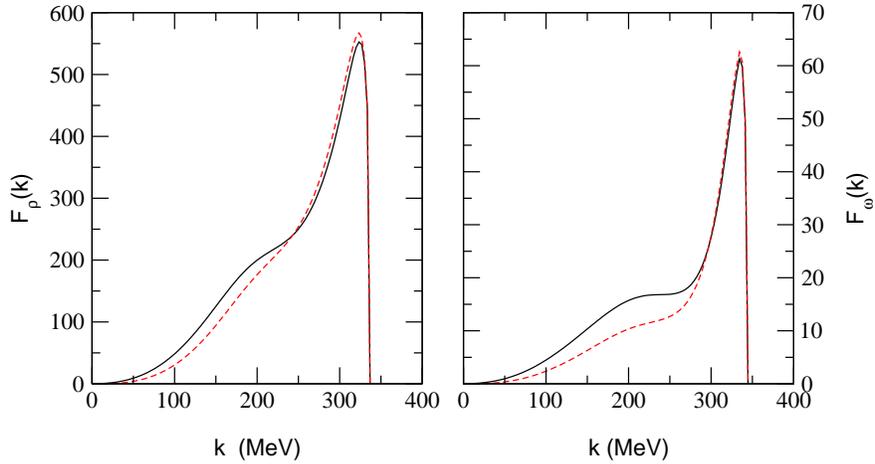}
\caption{The recoil photon spectrum $F_V^{}(k) = (384 \pi^3 M_V) \times
d\Gamma(V\to \pi^0\pi^0\gamma)/dk$ for $\rho \to \pi^0\pi^0\gamma$
(left panel) and $\omega \to \pi^0 \pi^0 \gamma$ (right panel) from
model (C). The solid lines are obtained with the $f_2$ meson contribution
while the dashed lines are without the $f_2$ meson contribution.}
\label{fig:res}
\end{figure}

The recoil photon spectrum functions $F_V^{}(k)$ are shown in
Fig.~\ref{fig:res} for the two decays in model (C).
Here the dashed lines are obtained without the $f_2$ meson and the solid
lines are with the $f_2$ meson.
This clearly shows that the contribution of the
$f_2$ meson is larger in the case of $\omega$ decay.
In the $\rho$ decay, the contribution of the $f_2$ meson is suppressed by
the $\sigma$ meson.
One can also find that the $f_2$ meson effect is sizable 
in the intermediate photon energies around $k \sim 200$ MeV.
Thus, precise measurements of the recoil photon spectrum for the $\omega$
decay would be very useful to identify the role of the $f_2$ meson.

In Table \ref{tab:res}, we also present the results without using the
tensor meson dominance, i.e., $G_{f\pi\pi} \neq G_{fVV}$.
Here, the coupling constant, $G_{fVV}$ is determined from the existing
experimental data for $f_2 \to \gamma\gamma$ decay \cite{PDG02}.
The values in the parentheses are obtained with $G_{fVV} = 3.12$.
Although the results are smaller than those with the tensor meson dominance,
we still see a sizable contribution from the $f_2$ meson.
Therefore, measuring $f_2 \to V\gamma$ decay is important not only to test
the tensor meson dominance but also to precisely constrain the role of the
$f_2$ meson in $V \to \pi^0 \pi^0 \gamma$ decay.

Finally we examine the sensitivity of $\Gamma(\omega \to \pi^0 \pi^0 \gamma)$
to several models for the $\rho$-$\omega$ mixing. 
We first note that the $\rho$-$\omega$ mixing amplitude may have momentum
dependence.
Following Ref. \cite{HHMK94}, 
we employ the momentum dependence of the mixing amplitude as
\begin{equation}
\mathcal{M}_{\rho\omega}^2(q^2) = \mathcal{M}_{\rho\omega}^2 \left[ 1 +
\lambda \left( \frac{q^2}{M_{\rm av}^2} - 1 \right) \right],
\label{q2M}
\end{equation}
where $M_{\rm av}$ is the average mass of the $\rho$ and $\omega$
mesons, and $\lambda \approx 1.5$.
Since the intermediate $\rho$ meson has the momentum with $q^2 < M_\rho^2$,
the momentum-dependence (\ref{q2M}) is expected to reduce the decay width.
Indeed, if we turn on the $q^2$ dependence of the mixing in model (C),
we obtain 
\begin{equation}
\mbox{BR}(\omega \to \pi^0 \pi^0 \gamma) = 6.13 \times 10^{-5},
\end{equation}
which is slightly lower than the result, $6.36 \times 10^{-5}$.
(See Table~\ref{tab:res}.)
The larger mixing amplitude $\mathcal{M}_{\rho\omega}^2 = -5000$ MeV$^2$
suggested by Ref. \cite{Scad97}, which is close to the QCD sum rule
prediction \cite{IJL96}, gives
\begin{equation}
\mbox{BR}(\omega \to \pi^0 \pi^0 \gamma) = 6.46 \times 10^{-5} \quad
(6.14 \times 10^{-5}),
\end{equation}
where the value in the parenthesis is obtained by allowing the $q^2$
dependence of the mixing.
Finally using the complex amplitude~\cite{GO98},
\begin{equation}
\mathcal{M}_{\rho\omega}^2 = (-3500 \pm 300 \mbox{ MeV}^2) + i (-300
\pm 300 \mbox{ MeV}^2),
\end{equation}
we also obtain the comparable results,
\begin{equation}
\mbox{BR}(\omega \to \pi^0 \pi^0 \gamma) = 6.34 \times 10^{-5} \quad
(6.12 \times 10^{-5}).
\end{equation}
Thus, the $\omega \to \pi^0 \pi^0 \gamma$ decay width does not
suffer from the model dependence on the $\rho$-$\omega$ mixing so much.
The $q^2$ dependence reduces the decay width by 5\% at most, which is
essentially the same for all the models.

\section{Summary}

In this paper, we have investigated the role of the mesons with higher mass
in the decays of $\rho \to \pi^0 \pi^0 \gamma$ and $\omega
\to \pi^0 \pi^0 \gamma$.
Among the spin-1 and spin-2 mesons that have not been considered so far, 
only the $f_2(1270)$ tensor meson
can give nontrivial contributions.
We have estimated its contribution {\em quantitatively\/} by using the
effective Lagrangian approach.
The effective Lagrangian we used was constructed by fully taking into
account the tensor structure of the $f_2$ meson interactions.
The coupling constants are determined from the experimental decay width
and assumption of tensor/vector meson dominance.

For the $\rho \to \pi^0 \pi^0 \gamma$ decay, the contribution from the
$f_2$ meson is relatively small.
This supports the previous conclusion that the $\sigma$ meson effect is
crucial in understanding the experimental data.
For the $\omega \to \pi^0 \pi^0 \gamma$ decay, however, we found 
a sizable contribution from the $f_2$ meson through the strong interference
with the other decay amplitudes, which can reproduce the experimentally
observed decay width.
In this case, the contribution from the $\sigma$ meson is small since it
can contribute only through the $\rho$-$\omega$ mixing.

In the viewpoint of the linear sigma model or unitarized chiral perturbation
theory, which are successful in explaining the $\rho \to \pi^0 \pi^0 \gamma$
decay \cite{BELN01,PHO02}, one does not have direct $V\sigma\gamma$
($V=\rho,\omega$) couplings.
Instead {\em effective\/} $V\sigma\gamma$ couplings are generated
by chiral loops for the $\rho\sigma\gamma$ coupling and by the
$\rho$-$\omega$ mixing for the $\omega\sigma\gamma$ coupling.
In the latter case the $\sigma$ meson couples to $\rho\gamma$
through a chiral loop, and then the $\rho$ converts into the $\omega$
by the $\rho$-$\omega$ mixing.
Although such approaches of provide consistent description for the
$V\sigma\gamma$ couplings, they underestimate the observed data for
$\Gamma(\omega \to \pi^0 \pi^0 \gamma)$.
This lead to a model where the $V\sigma\gamma$ couplings are treated in a
different footing, i.e, the direct $\omega\sigma\gamma$ coupling is allowed
\cite{GKY03} while the direct $\rho\sigma\gamma$ coupling is not
\cite{GSY03}.

Our results, however, show that an alternative explanation for the
$\omega \to \pi^0 \pi^0 \gamma$ decay can be provided without
introducing the direct $\omega\sigma\gamma$ coupling. 
By including the $f_2$ meson contribution, a consistent description
not only for $\Gamma(\rho \to \pi^0 \pi^0 \gamma)$ but also for
$\Gamma(\omega \to \pi^0 \pi^0 \gamma)$ could be obtained while keeping
the consistent treatment for the $V\sigma\gamma$ couplings.
This suggests that the direct $\omega\sigma\gamma$ coupling, if nonvanishing,
should be much smaller than the value estimated by Ref.~\cite{GKY03}.%
\footnote{Note that the analyses on $\omega$ photoproduction near threshold
\cite{OTL01} also show that it does not need a sizable contribution from the
$\sigma$ meson exchange.}
Therefore, careful analyses on the decay of $\omega \to \pi^+ \pi^-
\gamma$ both theoretically and experimentally would be interesting in
resolving the role of the $\sigma$ meson in the $\omega$ meson decays.
At present, only the upper limit of the branching ratio of the
$\omega \to \pi^+ \pi^- \gamma$ is known \cite{PDG02}.
We have also checked that our results are not sensitive to various models
on the $\rho$-$\omega$ mixing in the literature. 
Finally since the $f_2$ contribution can be seen in the recoil photon
spectrum of the $\omega \to \pi^0 \pi^0 \gamma$ decay around
$k \approx 200$ MeV, precise measurements of the photon spectrum should
confirm the $f_2$ meson contribution.

\acknowledgments

We are grateful to S. Capstick, S. Godfrey, T.-S. H. Lee, and E. Oset
for useful discussions.
This work was supported by Korea Research Foundation Grant
(KRF-2002-015-CP0074).

\end{document}